# Eavesdropping Risk Evaluation on Terahertz Wireless Channels in Atmospheric Turbulence

Yu Mei[1], Yongfeng Ma[1], Jianjun Ma[1], Lothar Moeller[2], John F. Federici[2]

[1]School of Information and Electronics, Beijing Institute of Technology, Beijing 100081, China

[2] Department of Physics, New Jersey Institute of Technology, 322 King Blvd., Newark, New Jersey 07102, USA

Corresponding author: Jianjun Ma (e-mail: jianjun_ma@bit.edu.cn).

This research was supported by National Natural Science Foundation of China (No. 6207106), U.S. National Science Foundation (No. ECCS-1102222), and Beijing Institute of Technology Research Fund Program for Young Scholars (No. 3050012222008).

**ABSTRACT** Wireless networks operating at terahertz (THz) frequencies have been proposed as a promising candidate to support the ever-increasing capacity demand, which cannot be satisfied with existing radio-frequency (RF) technology. Besides this, wireless channels in the THz range could be less vulnerable to interceptions because of their high beam directionality and small signal coverage. However, a risk for eavesdropping can still exist due to the multipath effects caused by unintended scattering when the channels operate in outdoor scenarios, such as in rain, snow, atmospheric turbulence, etc.

In this work, eavesdropping risks for THz channel passing atmospheric turbulences are evaluated from a physical layer perspective. Secrecy capacity and outage probability of a point-to-point THz wireless channel are derived by considering the multipath scattering effect. Deterministic and probabilistic eavesdropping attacks are assessed. Their dependence on turbulence strength, eavesdropper's position and channel conditions is investigated.

## I. INTRODUCTION

The evolving requirement for high data rate services and applications, such as online education and medical services, kiosk downloading and Internet of Things (IoT), has pushed research on THz wireless communications over several years [1]. Compared with its RF counterparts, THz wireless techniques can offer several advantages such as high data capacity, solution to communication blackout [2] and RF interference isolation. Its distinct feature, the requirement for line-of-sight (LOS) conditions for proper signaling, protects it against simple eavesdropping attacks [3]. In other words, THz wireless channels suffer lower multipath scattering and are less vulnerable to interceptions compared to RF channels because of the high beam directivity they are able to provide. However, in some outdoor scenarios (such as in rain, snow and atmospheric turbulence), where the LOS requirement can be scattered [4,5], eavesdropping of multipath components may take place [6].

Early studies of eavesdropping on THz signals have been performed in [7], which analyzed backscattered radiation of a LOS indoor channel. Subsequent to that work, methods employing multipath schemes [8] and multiplexed orbital angular momentum beams [9] were proposed and studied theoretically to overcome signal eavesdropping. However, only little is published about multipath induced eavesdropping risk of a THz link operating in adverse outdoor weather conditions, even though many efforts concentrating on FSO links have been discussed [10,11]. In this work, we try to investigate the possibility of eavesdropping on a THz channel propagating through atmospheric turbulence under consideration of turbulence strength, the eavesdropper's location and the channel conditions.

Section II presents the models for simulating signal attenuation and multipath component generation when a THz channel propagates in atmospheric turbulence. We also compare our models with experimental data. In Section III, we introduce models to evaluate deterministic and probabilistic eavesdropping risks on a THz channel due to multipath scattering . We conclude in Section IV our study with some remarks about security aspects of future THz links.

## II. SIGNAL ATTENUATION AND MULTIPATH SCATTERING IN ATMOSPHERIC TURBULENCE

TABLE I
DEFINITION OF KEY PARAMETERS

| Parameter | Definition |
|---|---|
| $\alpha_g$ | Total gaseous attenuation |
| $A_g$ | Absorption loss by water vapor |
| $S_g$ | Scattering loss |
| $G_F$ | Atmospheric attenuation |
| $G_D$ | Divergence loss |
| $G_{LOS}$ | LOS channel gain |
| $G_{NLOS}$ | NLOS channel gain |
| Cn2 | Refractive index structure parameter |
| $C_s$ | Secrecy capacity |
| $P_o$ | Outage probability |
| MSC | Maximum secrecy capacity |
| MOP | Minimum outage probability |

Wireless channels propagating in outdoor scenarios can be attenuated by gaseous absorption and scattered by water vapor, turbulences or bigger particles like rain drops [12]. The total gaseous attenuation is often written as $\alpha_g = A_g + S_g$ where $A_g$ stands for the absorption loss by water vapor and other gases (such as oxygen) [13,14]. $S_g$ is the scattering loss obtained by Rayleigh or Mie scattering theories [15]. However, when the channels passing atmospheric turbulence, there should be one more factor that attributes to the atmospheric attenuation $G_F$. This attenuation $A_t$ relates to the coefficient $\alpha_t$, which was first proposed by Naboulsi for electromagnetic waves propagating through weak turbulence [16] and can be expressed as $A_t = \alpha_t L = 2(23.17 k^{7/6} C_n^2 L^{11/6})^{1/2}$ [dB] [17] with $k = 2\pi/\lambda$ as the wave number, $L$ as the propagation distance in turbulence (equals link distance $d$) and $C_n^2$ as the refractive index structure parameter. $C_n^2$ classifies the turbulence strength as listed in Table 2 and is usually derived from the air velocity and/or temperature fluctuations [18]. For vertical or slant paths, $C_n^2$ can depend on the altitude as described in the Hufnagle-Valley Model [19]. However, Naboulsi's model is not designed for the moderate and strong turbulence conditions as experimentally described in [20] and [21]. Also, it does not consider aperture averaging on the receive side and is suitable for plane waves only [17]. Wilfert's method rates the attenuation [19] as

$$A_t = \alpha_t L = \left|10\log\left(1 - \sqrt{\sigma_I^2}\right)\right| \text{ [dB]} \quad (1)$$

for plane or spherical waves and considers averaging based on limited aperture size $D$. The term $\sigma_I^2$ is the scintillation index (i.e. normalized variance of irradiance) by Rytov approximations [22]. It could be used for predicting the propagation of infinite plane and spherical waves along a horizontal path in atmospheric turbulences over the whole turbulence strength. This equation holds only for turbulences with Rytov variances $\sigma_R^2 = 1.23 C_n^2 k^{7/6} L^{11/6}$ (for a plane wave) and $\beta_R^2 = 0.5 C_n^2 k^{7/6} L^{11/6}$ (for a spherical wave) being smaller than 1. This could always be satisfied at THz frequencies, whose much larger wavelength can isolate or reduce the influence of scintillation effects [4]. In the following we estimate the THz signal attenuation due to atmospheric turbulences by applying Eq. (1) and write the atmospheric attenuation as

$$G_F = \exp\left[-(\alpha_t + \alpha_g)d\right] = \exp\left[-\alpha_{atm} d\right] \quad (2)$$

Atmospheric turbulence is caused by spatial and temporal temperature/pressure inhomogeneities in air [23,24] and can usually be modelled as a large number of air pockets with varying sizes (between a small scale size $l_0$ and a large scale size $L_0$), temperatures and pressures, which could also lead to beam divergence. The signal loss caused by divergence can be obtained as

$$G_D = 4A/\left(\pi d^2 \alpha_A^2\right) \quad (3)$$

with $A$ being the effective receiving area of Bob's antenna and $\alpha_A$ being the full divergence angle of the beam.

Combining both, the atmospheric attenuation (Eq. (2)) and divergence attenuation (Eq. (3)), yields the total loss of the LOS channel

$$G_{LOS} = G_F G_D = \frac{4A e^{-\alpha_{atm} d}}{\pi d^2 \alpha_A^2} \quad (4)$$

Turbulence induced signal variation can be split into a slow component and a fast component [4]. The former one is an averaged value ($\alpha_t$) caused by variation of refractive index. We refer to the channel loss given by Eq. (4) as deterministic attenuation for our analysis of deterministic eavesdropping in section 3. The latter one is due to the fast and random fluctuation of the refractive index, which accounts for the probabilistic eavesdropping in Section 4.

TABLE II
CLASSIFICATION OF TURBULENCE STRENGTH [22]

| Turbulence strength | $C_n^2$ (m$^{-2/3}$) |
|---|---|
| Weak | < 10$^{-17}$ |
| Moderate | (10$^{-17}$, 10$^{-13}$) |
| Strong | > 10$^{-13}$ |

Our model assumes a point-to-point outdoor THz wireless channel with the configuration shown in Fig.1(a). A transmitter (Alice) sends information to a legitimate receiver (Bob) by a LOS channel through absorbing and scattering turbulences, which is described by Eq. (4). An eavesdropper (Eve) outside the beam but positioned in its nearby proximity aims to capture data through a NLOS path. To ensure a conservative risk evaluation, we assume the eavesdropper has complete knowledge of the legitimate channel's parameters and has sufficient computational capabilities. The positions of Alice and Bob are fixed. Eve adjusts its antenna position and steering direction for optimal detection of the captured signal, which can be described by the deterministic channel gain of the NLOS path. In this work, we apply the widely used *single-scattering model* for the calculation of this term [25,26]. When the signal (Fig. 1(a)) is transmitted along the $x$-axis from Alice at (0,0) to Bob at (d,0) with Eve at (x,y) whose NLOS channel gain $G_{NLOS}$ [27] reads

$$G_{NLOS} = \int_{L_a}^{L_b} \Omega(l) p(\mu) \alpha_{atm} e^{-\alpha_{atm}[l+\sqrt{(x-l)^2+y^2}]} dl \quad (5)$$

The limits for $l$ are expressed by an upper and lower bound ($L_a$, $L_b$), which describes the scattering region. $\Omega(l)$ denotes the solid angle from the receiving area to the scattering center as

$$\Omega(l) = \frac{A}{[(x-l)^2+y^2]^{3/2}} \frac{(x-l)+y\tan\alpha}{\sqrt{1+\tan^2\alpha}} \quad (6)$$

The factor $p(\mu)$ is defined as scattering phase function indicating the probability distribution of scattering angle. When a generalized Henyey-Greenstein function is adopted, it reads

$$p(\mu) = \frac{1-g^2}{4\pi}\left[\frac{1}{(1+g^2-2g\mu)^{3/2}} + f\frac{3\mu^2-1}{2(1+g^2)^{3/2}}\right] \quad (7)$$

with $\mu = (x-l)/[(x-l)^2+y^2]^{1/2}$ representing the cosine of scattering angle and $g$ is an asymmetry factor related to wavelength, scattering particle radius and refractive index [28].

To evaluate the accuracy of the model, we have conducted measurements by employing a 625 GHz wireless channel propagating through emulated atmospheric turbulences using a weather chamber [4]. The atmospheric turbulence was generated by introducing air flows at different temperatures (35°, 55° and 70°) and air speeds (28.6 m/s and 41.6 m/s) into a weather chamber. The turbulence strength can be adjusted from $3.5\times10^{-11}$ m$^{-2/3}$ to $2.3\times10^{-9}$ m$^{-2/3}$, which corresponds to maximum Rytov variances of $\sigma_R^2 = 0.059$ and $\beta_R^2 = 0.037$ for a plane wave and spherical wave, respectively, i.e. making Eq. (1) applicable for our applications.

Two Teflon lenses with 32 mm focal length and 5 cm diameter collimate our THz beam which can be well represented by a plane wave. Antenna gains as high as 55 dBi [29] have been obtained with parabolic offset reflectors whereas our lenses provide about 20 dBi but benefit from relatively easy handling. Fig. 2(b) shows theoretical results which qualitatively agree with the experimental data taken over a channel distance of 1m and confirm the applicability of Eq. (4).

The attenuation across the THz spectrum obtained with this model is plotted for turbulence strengths from $C_n^2 = 3.5\times10^{-11}$ m$^{-2/3}$ to $2.3\times10^{-9}$ m$^{-2/3}$ in Fig. 2(c). The attenuation caused by turbulences follows the spectral absorption in an undistorted path but is offset. Wireless channels operating at frequency windows around 140, 220 and 340 GHz, have been utilized to demonstrate long distance signaling [30-32]. The transmission band at 675 GHz has been proposed as most suitable candidate for practical realization of 1 Tbps data transmission [33]. However, we choose for our modelling wireless channels propagating over 1km and operating at 140, 220, 340 and 675 GHz to predict their vulnerability to eavesdropping.

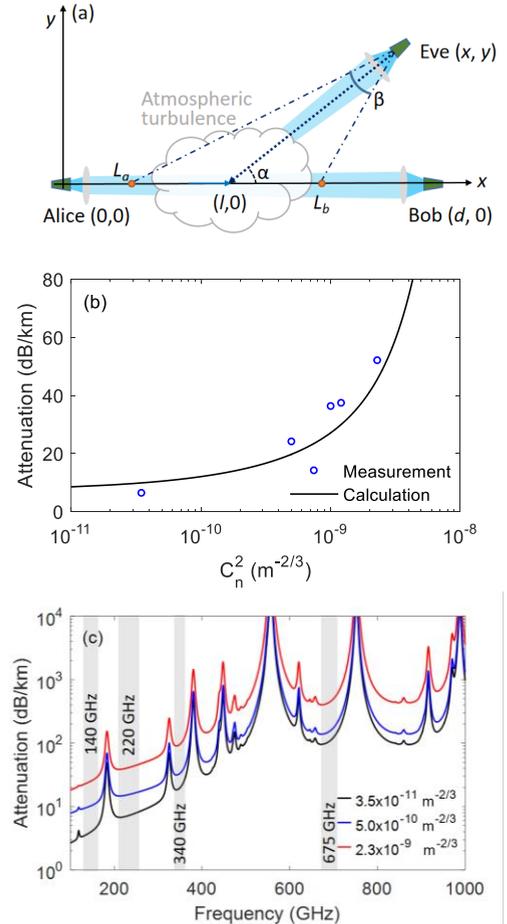

**FIGURE 1.** (a) Geographic of a point-to-point THz channel with an eavesdropping attacker (Eve) located outside of the channel path with positions of Alice and Bob always fixed. (α is the scattering angel in direction of Eve); (b) Comparison of measured data with predicted deterministic attenuation for a wireless channel at 625 GHz propagating through emulated atmospheric turbulences. Hot and dry air was introduced into a weather chamber to generate atmospheric turbulences with a measured relative humidity of RH = 10%; (c) Attenuation due to atmospheric turbulence with different strengths. (pressure $P$ = 1013 hPa, humidity RH = 20%, channel distance $d$ =1km).

### III. EAVESDROPPING RISK EVALUATION AND DISCUSSIONS

#### A. DETERMINISTIC EAVESDROPPING RISK

Table 3 shows the basic system parameters of our channel model that assumes equal receiver sensitivity on Bob's and Eve's sides. The beam divergence $\alpha_A$ = 20 mrad accounts for small Tx misalignments at THz frequencies. Fig. 2(a) shows the channel gain at turbulence strengths up to $C_n^2 = 1.0 \times 10^{-10}$ m$^{-2/3}$, which corresponds to $\beta_0^2 = 0.49$. Solid black lines represent the evolution of the channel gain $G_{LOS}$ for the LOS channel to Bob as a function of the turbulence strength $C_n^2$ and the dashed blue lines stand for the NLOS channel gain received by Eve. In practice, if Eve is close to the LOS path and covered by the LOS beam, it could obtain the NLOS and LOS components together. But in this work, we neglect the LOS beam coverage (which was discussed in [7]) and just consider a point-to-point channel.

For weak turbulences in Fig. 2(a), Bob's received power becomes stronger while Eve's received power stays about constant. When the turbulence strength reaches about $C_n^2 = 5.8 \times 10^{-11}$ m$^{-2/3}$, the two lines cross over and a significant eavesdropping risk comes. Eqs. (4) and (5) suggest that successful eavesdropping becomes more difficult after increasing the minimum distance from Eve to the LOS channel path.

Wyner's secrecy capacity metric [34] is applied to evaluate deterministic eavesdropping risk faced by wireless systems. It was defined as the highest data rate that can be attained from Alice to Bob with keeping Eve ignorant [35] and can be expressed as

$$C_s = \left[ I(X;Y) - I(X;Z) \right]^+ \quad (8)$$

with $[x]^+ = \max\{0, x\}$ indicating that the value will be 0 when $x \leq 0$ and will be $x$ when $x > 0$. Parameters $X$, $Y$ and $Z$ represent the signals of the Alice, Bob, and Eve, respectively. $I(X;Y)$ and $I(X;Z)$ denote the mutual information of LOS and NLOS channels [34], respectively. The expressions for them could be found in [6] when an on-off keying (OOK) modulation format with a duty cycle $q$ and a Poisson distribution of photoelectrons is assumed [27]. They could be expressed as

$$I(X;Y) = q(\lambda_L + \lambda_b)\log(\lambda_L + \lambda_b) + \lambda_b \log(\lambda_b) \\ - (q\lambda_L + \lambda_b)\log(q\lambda_L + \lambda_b) \quad (9\text{-}1)$$

and

$$I(X;Z) = q(\lambda_N + \lambda_e)\log(\lambda_N + \lambda_e) + \lambda_e \log(\lambda_e) \\ - (q\lambda_N + \lambda_e)\log(q\lambda_N + \lambda_e) \quad (9\text{-}2)$$

OOK modulation is relatively easy to implement in lab test beds and is applied here, although several other higher order schemes (such as QPSK, QAM) have been demonstrated [36-38]. In the calculation for $I(X;Y)$ and $I(X;Z)$, we set $\lambda_L = \tau\eta G_{LOS} P/E_p$ and $\lambda_N = \tau\eta G_{NLOS} P/E_p$ representing the mean numbers of detected photoelectrons of signal component in each bit slot for the LOS and NLOS channels, respectively. $P$ is the output power from Alice and $\eta$ is the receiver efficiency, which are identical for Bob and Eve in the modelling, which is available as in [39]. $E_p$ is the energy per THz photon and $\tau$ is an integration time of the receivers. $\lambda_b$ and $\lambda_e$ represent the mean number of detected photoelectrons of a background radiation component in each bit slot. The THz radiation is converted to direct current (DC) using a rectifying diode connected to the output of the antenna. We take the photoelectron in consideration in our calculation and the signal-to-noise ratio (SNR) of a receiver can be obtained by dividing $\lambda_L$ by $\lambda_b$ or $\lambda_N$ by $\lambda_e$. It is noteworthy that since the LOS and NLOS channel gains obtained by Eqs. (4) and (5) are averaged (deterministic) values, the eavesdropping risk predicted by Eq. (8) can be considered to be deterministic.

TABLE III
LIST OF PARAMETERS USED IN CALCULATIONS

| Parameter | Value |
| --- | --- |
| Location of Alice | (0m, 0m) |
| Location of Bob | (1km, 0m) |
| Location of Eve | (750m, 30m) |
| Temperature ($T$) | 30 °C |
| Pressure ($Pr$) | 1013 hPa |
| Relative humidity (RH) | 80% |
| divergence angle ($\alpha_A$) | 20 mrad |
| Turbulence strength ($C_n^2$) | $5.8 \times 10^{-11}$ m$^{-2/3}$ |
| Output power of Alice ($P$) | 10 mW [39] |
| Receiver efficiency ($\eta$) at Bob and Eve | 0.1 |
| Full angle of field of sight (FOV) | 10° |
| Receiving aperture diameter ($D$) | 5cm for Bob<br>5cm for Eve |
| Receiver sensitivity by SNR | 0dB for Bob<br>0dB for Eve |
| Intended data rate ($R$) | 10 Gbps |

The secrecy capacity with respect to an arbitrary position of Eve is shown in Fig. 2(b) and (c). In Fig. 2(b), its $x$-position varies from 0 m to 1 km while $y = 30$ m. When Eve is located with its $x$-position in the range of [28m, 742m], the channel secrecy capacity would be 0 Gbps. This means no secure data transmission could be achieved and we call this area as 'insecure region'. If we set $x = 750$m and change its $y$-position, the evolution of the secrecy capacity is plotted in Fig. 2(c). At positions of $y \leq 30$m, there is 'insecure region'. When $y > 30$m, the secrecy capacity increases dramatically and reaches to a maximum secrecy capacity (MSC). This indicates that the method - increasing the minimum distance from Eve to the LOS channel path, is capable to reduce and overcome eavesdropping risk caused by atmospheric turbulence.

The secrecy capacity distribution with respect to arbitrary 2-D positions of Eve is plotted in Fig. 2(d) with its color bar denotes the secrecy capacity in Gbps. The brightest color represents the MSR of 44 Gbps and the dark blue region represents $C_s = 0$ Gbps, which is the insecure region. The horizontal and vertical white lines stand for the evolution of secrecy capacity versus *x*- and *y*- positions of Eve as plotted in Fig. 2(b) and (c), respectively.

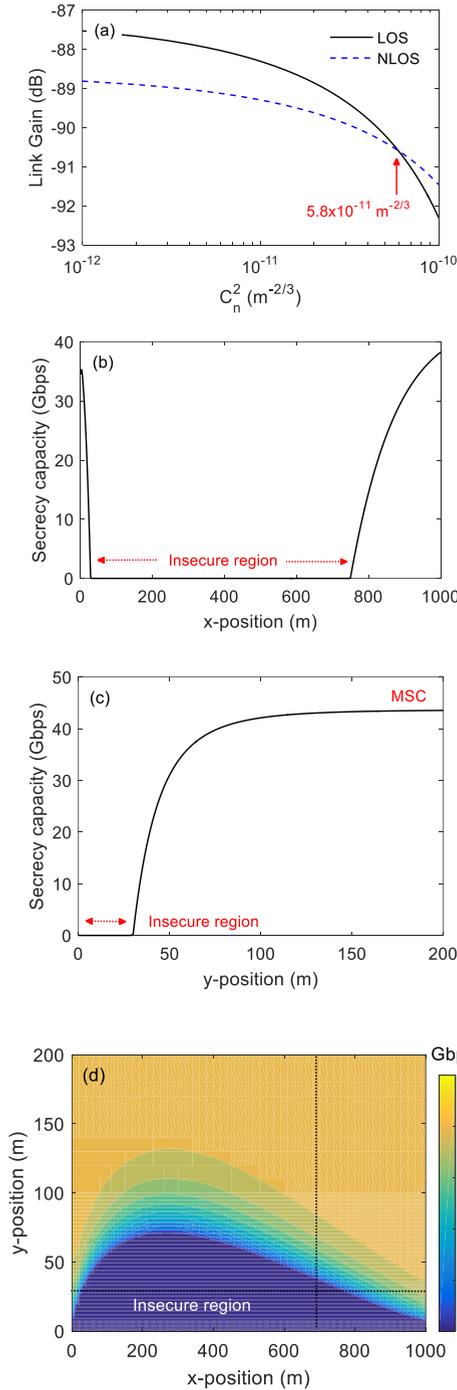

FIGURE 2. (a) Evolution of channel gain received by Bob (LOS) and Eve (NLOS) versus turbulence strength when Eve is located at (750m, 30m); (b) Evolution of channel secrecy capacity distribution versus *x*-position of Eve (*y* = 30m); (c) Evolution of channel secrecy capacity distribution versus *y*-position of Eve (*x* = 750m); (d) Secrecy capacity distribution for 2-D positions of Eve with a unit of Gbps in the color bar.

To see the dependence of eavesdropping risk on carrier frequencies, we calculate the secrecy capacity of the channel operating at 140, 220, 340 and 675 GHz as in Fig. 3(a). With the increasing of carrier frequencies, the MSC value decreases significantly and the insecure region is extended due to more serious scattering and higher gaseous attenuation suffered by higher frequencies as in Fig. 1(c). At 675 GHz, the secrecy capacity reaches to $C_s = 0$ Gbps, which means the whole region would be insecure. This is also indicated in the inserted plot in Fig. 3(a) with a whole region as dark blue. Thus, wireless channels operating at higher carrier frequencies would more serious multipath scattering and eavesdropping risks.

Fig. 3(b) shows the eavesdropping response of a 340 GHz channel on turbulence strength variation. The MSC is decreased and the insecure region is expanded when the turbulence strength increases from $C_n^2 = 10^{-12}$ m$^{-2/3}$ to $10^{-10}$ m$^{-2/3}$. This trend is consistent with the calculation in Fig. 2(a), where stronger atmospheric turbulences would decrease the difference between Bob's and Eve's receive power and lead to more serious eavesdropping risk.

The atmospheric turbulence would also lead to beam divergence and pointing errors. In Fig. 3(c), we calculate the variation of secrecy capacity with respect to the change of divergence angle. When $\alpha_A$ increases from 25 mrad to 35 mrad, the insecure region (the region for $C_s = 0$ Gbps) is expanded and the MSC is reduced from $C_s = 27$ Gbps to 13 Gbps.

Since we assume Eve has the complete channel state information (CSI) of the legitimate channel and has sufficient computational capabilities, it could maximize the captured information by increasing its receiver sensitivity and field-of-view (FOV) full angle. In Fig. 3(d), we estimate the secrecy capacity for different SNR values and different FOV angles of Eve. The insecure region (the region for $C_s = 0$ Gbps) is enlarged obviously with Eve's SNR decreasing from 6 dB to 0 dB, even though the MSC is not affected. That's because the MSC only depends on the receiver sensitivity of Bob. However, the change of capacity is not so pronounced when Eve's FOV angle is 5° or 20°, which means that this would not be a good strategy for reduction eavesdropping risks.

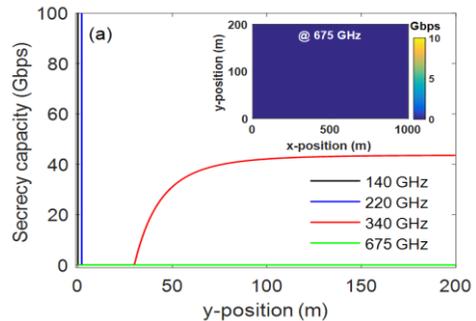

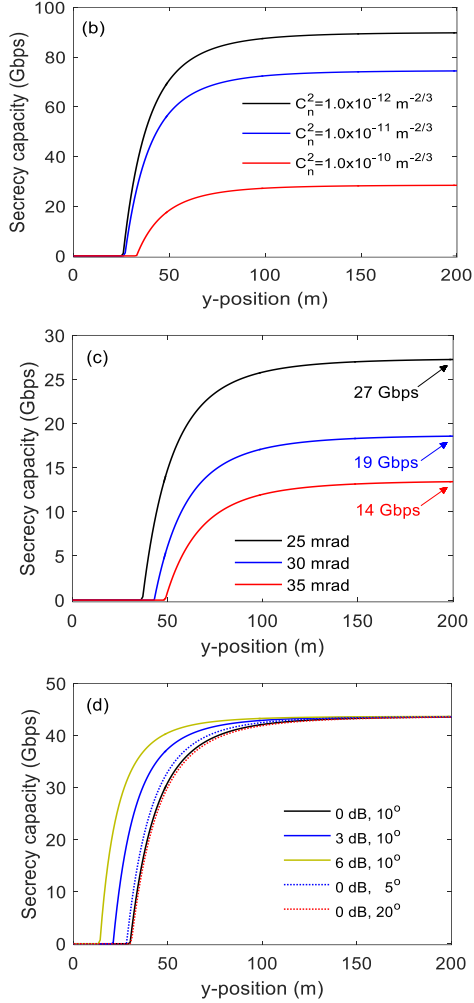

**FIGURE 3.** Variation of secrecy capacity with respect to the *y*-position of Eve under different (a) carrier frequencies, (b) turbulence strengths, (c) divergence angles, and (d) receiver sensitivities and FOV angle for Eve. Inset of (a): secrecy capacity distribution for 2-D positions of Eve with a unit of Gbps in the color bar.

## B. PROBABILISTIC EAVESDROPPING RISK

Since the received power and channel gain should always fluctuate spatially and temporally due to the scattering induced scintillation effects along the channel path [40], a random variable, i.e. outage probability, should be employed to assess the eavesdropping risk. The outage probability is defined as the probability that the instantaneous secrecy capacity falls below a target rate $R$ and it can be obtained by $P_0(R) = P_r\{C_s < R\}$ [42] with $R \geq 0$ always. This expression can be rewritten for our model as

$$P_o(R) = \int_{C_s \leq R} f_{LOS}(G_{LOS}) dG_{LOS} = \int_0^G f_{LOS}(G_{LOS}) dG_{LOS} \quad (10)$$

with $G$ as the solution of $C_s = R$. $G_{LOS}$ becomes an instantaneous LOS channel gain here. The probability density function of the instantaneous LOS channel gain can then be expressed [22] as

$$f_{LOS}(G_{LOS}) = \frac{1}{G_{LOS}\sqrt{2\pi\sigma_r^2}} \cdot$$
$$\exp\left[-\frac{\left(\log(G_{LOS}/\overline{G_{LOS}}) - \langle\log(G_{LOS}/\overline{G_{LOS}})\rangle\right)^2}{2\sigma_r^2}\right] \quad (11)$$

Here, the log(·) term means the log-normal model is used to describe the signal distribution in atmospheric turbulence with the Raytov variance $\sigma_R^2 < 1$ or $\beta_R^2 < 1$, which is usually regarded as a good tool [41]. $\overline{G_{LOS}}$ is the mean value of the random variable $G_{LOS}$. $\langle\log(G_{LOS}/\overline{G_{LOS}})\rangle$ stands for the mean value of $\log(G_{LOS}/\overline{G_{LOS}})$. The parameter $\sigma_r^2$ represents the variance of $\log(G_{LOS})$. In the presence of atmosphere turbulence, $\sigma_r^2$ is defined as the Rytov variance characterizing the strength of turbulence over a transmission channel for a spherical wave. The variance $\sigma_r^2 = 0.5 C_n^2 k^{7/6} d^{11/6}$ is a function of atmosphere refraction structure parameter $C_n^2$, the wave number $k$, and the channel distance $d$.

We conduct the outage probability by Eq. (10) and show the results in Fig. 4 with the basic parameter values as in Table 3. We set the intended data rate $R$= 10 Gbps which is a common achievement for wireless channels with carriers at THz range [43]. Identical to the trend in Fig. 3(a), the outage probability is very sensitive to the carrier frequency as shown in Fig. 4(a). The 140 GHz and 220 GHz channels are almost secure over the whole region with minimum outage probability (MOP) below $10^{-4}$, which means there is no insecure region (defined as the region with $P_o = 1$). Oppositely, the 675 GHz channel definitely suffers serious risk because its MOP always 1. For the 340 GHz channel, its outage probability starts to change when $y$ =33m and then decreases significantly to a constant value of MOP = 0.16%. Therefore, in the turbulence regime, higher carrier frequencies would suffer more signal loss, multipath scattering enlarge, and finally higher eavesdropping risk.

Fig. 4(b) presents the variation of the outage probability for a 340 GHz channel propagating through atmospheric turbulence at different strengths. It is shown that, due to the turbulence induced channel degradation, the outage probability increases significantly when the turbulence becomes stronger and the eavesdropping risk becomes more serious.

In Fig. 4(c), the impact of the beam divergence and pointing error are shown. Identical trend is observed and the same conclusion could be obtained as in Fig. 3(c).

Fig. 4(d) shows the impact of Eve's receiver sensitivity and FOV angle. When Eve decrease its receiver's BER value from 6dB to 0 dB, obvious expansion on insecure region (the region for $P_o = 1$) could be observed. This means Alice and Bob could utilize noise properties to reduce the SNR at Eve's side and minimize or avoid an eavesdropping risk. However, the response of insecure region on Eve's FOV angle is not so evident, which indicates that variation of FOV angle would not be a good strategy for the reduction of eavesdropping risk.

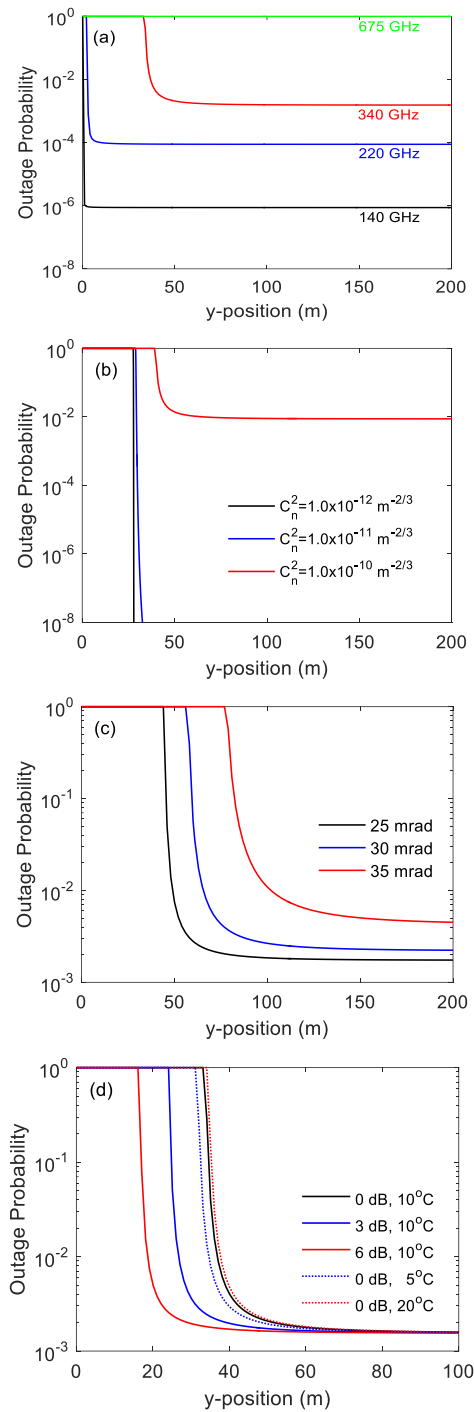

**FIGURE 4.** Variation of outage probability with respect to the y-position of Eve for different (a) carrier frequencies, (b) turbulence strengths, (c) divergence angles, and (d) receiver sensitivities and FOV angle of Eve.

## IV. CONCLUSIONS

In this paper, we evaluate the potential of eavesdropping risk on a point-to-point THz wireless channel passing through atmospheric turbulence when an unauthorized user (eavesdropper) locates in nearby proximity and tries to capture confidential information. A theoretical model combining signal attenuation due to turbulence, gaseous absorption and beam divergence is proposed to estimate the multipath components caused by turbulence-induced scattering effect. Deterministic and probabilistic eavesdropping attacks are evaluated by considering the influence of turbulence characteristics and channel conditions. It has been shown that wireless data transmission with lower multipath scattering and eavesdropping risk could be achieved by reducing the carrier frequencies, increasing the minimum distance from Eve to the LOS channel path and introducing random noise. We present a comprehensive model for estimation the secrecy performance of a THz channel. Its implementations can minimize an eavesdropping risk on the physical layer.

Physical layer techniques, such as cooperative nodes, noise randomness and multi-antenna techniques, are usually developed to reduce eavesdropping risk from modern wireless communication systems [44]. Our model and findings can be useful helpful for testing new methods to minimize risk for eavesdropping.

Technology, Newark, NJ, USA and in 2015, under the guidance of Prof. John F. Federici. His Ph.D. dissertation was about the weather impacts on outdoor terahertz and infrared wireless communication links.

In 2016, he joined Prof. Daniel M. Mittleman's group at Brown University, RI, USA, as a postdoctoral research associate. In 2019, he joined Beijing Institute of Technology, Beijing, China, as a professor. His current research interests include terahertz and infrared wireless communications, terahertz time domain detection, terahertz waveguides, terahertz indoor/outdoor propagation and link security.

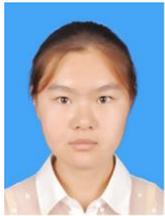

**Yu Mei** was born in Jiamusi, China, in 1996. She received the bachelor degree in Information Countermeasure Technology from Beijing Institute of Technology, China in 2018. From 2019, she studies as a master student in Communication Technology in Beijing Institute of Technology. Her main research direction is terahertz indoor/outdoor propagation and link security.

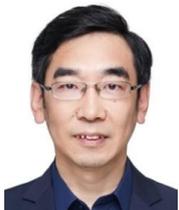

**Yongfeng Ma** was born in Henan province, China, in 1975. He received the master's degree in test and measurement technology from Harbin University of Science and Technology and the Ph.D. degree in communication and information system from Beijing Institute of Technology, in 2001 and 2004, respectively.

In 2004, he joined Beijing Institute of Technology and has been involved in research and development activities in high speed signal acquisition and processing since then. His current research interests include wireless communications based on backscatter, integration of laser communication and ranging.

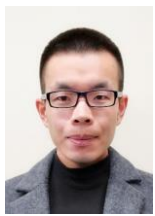

**Jianjun Ma** was born in Qingdao, China, in 1986. He received the Ph.D. degree in applied physics from New Jersey Institute of

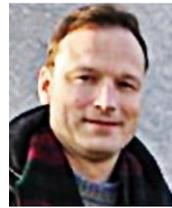

**Lothar Moeller** received the Dipl. Phys. degree in Physics and Dipl. Ing. Degree in Electrical Engineering from RWTH Aachen, Germany in 1992 and 1993, respectively and a Ph.D. in Electrical Engineering from ETH Zurich, Switzerland in 1996. He joined Bell Labs in 1997 where he researched and developed novel modulation and equalization techniques for a variety of technology fields including optical fiber transmission, THz communication systems with ultra-high capacity, THz security scanners, and detection of cell phone radiation using MRI. Since 2014 Dr. Moeller is with the R&D division of TE Connectivity where he focuses on undersea telecommunication systems. He became a research professor at the New Jersey Institute of Technology and a guest professor at the Shanghai Jiao Tong University in 2014 and 2005, respectively.

**John F. Federici** received his B.S. degree in Physics from the University of Notre Dame and his Ph.D. degree in Plasma Physics

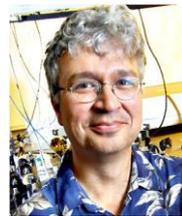

from Princeton University in 1983 and 1989, respectively. After a post-doctoral appointment at Bell Laboratories, he joined the Physics Department faculty at the New Jersey Institute of Technology. Since his days at Bell Laboratories, he has been active in THz spectroscopy, imaging, and most recently THz wireless communications research.